\documentclass[manuscript]{aastex}
\usepackage{epsfig}
\usepackage{amsmath}
\begin{document}

\title{High-Accuracy Quartic Force Field Calculations for the Spectroscopic
Constants and Vibrational Frequencies of $1 ^1A'$ $l$-C$_3$H$^-$: A Possible Link
to Lines Observed in the Horsehead Nebula PDR}

\date{\today}
\author{Ryan C. Fortenberry}
\affil{NASA Ames Research Center, Moffett Field, CA 94035-1000, U.S.A.}
\author{Xinchuan Huang}
\affil{SETI Institute, 189 Bernardo Avenue, Suite 100, Mountain View, CA 94043, 
U.S.A.}
\author{T. Daniel Crawford}
\affil{Department of Chemistry, Virginia Tech, Blacksburg, VA 24061, U.S.A.}
\author{Timothy J. Lee}
\email{Timothy.J.Lee@nasa.gov}
\affil{NASA Ames Research Center, Moffett Field, CA 94035-1000, U.S.A.}

\begin{abstract}

It has been shown that rotational lines observed in the Horsehead nebula PDR
are probably not caused by $l$-C$_3$H$^+$, as was originally suggested.  In the
search for viable alternative candidate carriers, quartic force fields are
employed here to provide highly accurate rotational constants, as well as
fundamental vibrational frequencies, for another candidate carrier: $1\ ^1A'$
C$_3$H$^-$.  The $ab\ initio$ computed spectroscopic constants provided in this
work are, compared to those necessary to define the observed lines, as accurate
as the computed spectroscopic constants for many of the known interstellar
anions.  Additionally, the computed $D_{eff}$ for C$_3$H$^-$ is three times
closer to the $D$ deduced from the observed Horsehead nebula lines relative to
$l$-C$_3$H$^+$.  As a result, $1\ ^1A'$ C$_3$H$^-$ is a more viable candidate
for these observed rotational transitions.  It has been previously proposed
that at least C$_6$H$^-$ may be present in the Horsehead nebular PDR formed by
way of radiative attachment through its dipole-bound excited state.  C$_3$H$^-$
could form in a similar way through its dipole-bound state, but its valence
excited state increases the number of relaxation pathways possible to reach the
ground electronic state.  In turn, the rate of formation for C$_3$H$^-$ could
be greater than the rate of its destruction.  C$_3$H$^-$ would be the seventh
confirmed interstellar anion detected within the past decade and the first
C$_n$H$^-$ molecular anion with an odd $n$.


\end{abstract}

\maketitle

{\bf{Keywords:}} astrochemistry $-$ ISM: individual objects: Horsehead nebula $-$
ISM: lines and bands $-$ ISM: molecules $-$ molecular data $-$ radio lines: ISM

\section{Introduction}

Recent work by \cite{Huang13C3H+} has questioned the attribution of lines
observed in the Horsehead nebula photodissociation region (PDR) to
$l$-C$_3$H$^+$.  Quartic force fields (QFFs) computed from high-level $ab\
initio$ quantum mechanical energies analyzed using perturbation theory at
second order \citep{Papousek82} are known to produce highly accurate
spectroscopic constants.  Even though the $B_0$ computed by \cite{Huang13C3H+}
(11 262.68 MHz) is within 0.16\% of the $B$-type rotational constant derived
from the observations by \cite{Pety12} (11 244.9474 MHz), the computed $D_e$ of
4.248 kHz differs by 44.5\% from the observed $D$ value of 7.652 kHz.  This
``error" is more than an order of magnitude larger than any other error for a
computed $D_e$ of a cation (using similar levels of theory) as compared to
known high-resolution experimental data.  Furthermore, the sextic distortion
constant, $H$, differs by three orders of magnitude.  As a result, it is
unlikely that $l$-C$_3$H$^+$ corresponds to the lines observed by
\cite{Pety12}.

This result motivates the question, ``What is the carrier of these lines?"  If
these observed lines are, in fact, related to one another, certain inferences
can be made about the molecular carrier.  To match the rotational constants
derived from the transition energies corresponding to the observed lines, the
carrier is either linear or quasi-linear, almost certainly composed of three
carbon atoms as well as a single hydrogen atom, and closed-shell since there
are no splittings in the lines as required for the rotational spectra of
open-shell molecules \citep{McCarthy13comm}.  All of these criteria are, in
fact, met by $l$-C$_3$H$^+$, but this cation's difference between observational
and high-accuracy theoretical rotational constants, especially the $D$
constant, discussed above and by \cite{Huang13C3H+}, probably rules it out.  As
a result, the quasi-linear anion, $1\ ^1 A'$ $l$-C$_3$H$^-$, remains as the
most likely candidate carrier of the Horsehead nebula PDR rotational lines of
interest especially since anions have been shown to be more abundant in the
interstellar medium (ISM) than originally thought \citep{Cordiner13}, and there
has been reason to suspect the presence of C$_6$H$^-$ in the Horsehead nebula
PDR \citep{Agundez08}.

Even though the most stable singlet isomer of C$_3$H$^-$ is the cyclic form,
$c$-C$_3$H$^-$, the barrier to isomerization is high enough ($\> 45$ kcal
mol$^{-1}$) for the quasi-linear $C_s$ isomer to be kinetically stable
\citep{Lakin01}.  Various mechanisms for interstellar synthesis of this anion
are possible \citep{Millar07, Herbst08, Larsson12, Senent13} and are probably
related to those responsible for the creation of the related C$_{2n}$H$^-$ for
$n=2-4$ anions previously detected in the ISM \citep{McCarthy06, Cernicharo07,
Brunken07C8H-}.  Furthermore, radical C$_3$H in both the linear and cyclic
forms has also been detected in the ISM \citep{Thaddeus85, Yamamoto87cC3H}
suggesting the possible interstellar existence of the anion.

Additionally, C$_3$H$^-$ is of astronomical interest since it has been
computationally shown by \cite{Fortenberry133DBS} to possess not only a rare
dipole-bound singlet excited electronic state (the $2\ ^1A'$ state) but also an
even more rare valence excited state ($1\ ^1A''$) below the electron binding or
electron detachment energy.  In fact, the valence electronically excited state
is the only such state thus far proposed to exist for an anion of this size
which also contains only first-row atoms \citep{Fortenberry11dbs,
Fortenberry112dbs, Fortenberry133DBS}.  The valence excited state and the bent
structure of C$_3$H$^-$ are both the result of an unfilled $\pi$ orbital.  The
two components of the HOMO $\pi$-type orbital split when the additional
electron in the anion spin-pairs with the lone electron in the radical's
$\pi$-type HOMO.  A carbene and bent structure are thus created.  The valence
($1\ ^1A''$) state of C$_3$H$^-$ is then the product of an excitation from the
occupied portion of the split $\pi$ orbital into the unoccupied portion, an
uncommon process not present in the C$_{2n}$H$^-$ anions.  Furthermore, anions
have been proposed as carriers of some diffuse interstellar bands (DIBs)
\citep{Sarre00, Cordiner07, Fortenberry13CH2CN-}, and the two electronically
excited states of C$_3$H$^-$ may be of importance to the DIBs and to the
chemistry of PDRs, as well.

\section{Computational Details}

The spectroscopic constants and fundamental vibrational frequencies of $1\
^1A'$ $l$-C$_3$H$^-$ are computed through the established means of QFFs
\citep{Huang08}.  Starting from a restricted Hartree-Fock (RHF)
\citep{ScheinerRHF87} coupled cluster \citep{Lee95Accu, Shavitt09, ccreview}
singles, doubles, and perturbative triples [CCSD(T)] \citep{Rag89} aug-cc-pV5Z
\citep{Dunning89, aug-cc-pVXZ, Dunning01} geometry further corrected for core
correlation effects from the Martin-Taylor (MT) basis set \citep{Martin94}, a
grid of 743 symmetry-unique points is generated.  Simple-internal coordinates
for the bond lengths and $\angle$ H$-$C$-$C are coupled to linear LINX and LINY
\citep{intder} coordinates exactly as those defined in
\cite{Fortenberry12hococat} for HOCO$^+$.  Displacements of 0.005 \AA\ for the
bond lengths, 0.005 rad for the bond angle, and 0.005 for the LINX and LINY
coordinates and the associated energies computed at each point define the QFF,
which is of the form:
\begin{equation}
V=\frac{1}{2}\sum_{ij}F_{ij}\Delta_i\Delta_j +
\frac{1}{6}\sum_{ijk}F_{ikj}\Delta_i\Delta_j\Delta_k +
\frac{1}{24}\sum_{ijkl}F_{ikjl}\Delta_i\Delta_j\Delta_k\Delta_l,
\label{VVib}
\end{equation}
where $\Delta_i$ are the displacements and $F_{ij\ldots}$ are force constants
\citep{Huang08}.

At each point, CCSD(T)/aug-cc-pVXZ (where $X=T,Q,5$) energies are computed and
extrapolated to the complete basis set (CBS) limit via a three-point formula
\citep{Martin96}.  Additionally, energy corrections are made to the CBS energy
for core correlation and for scalar relativistic effects \citep{Douglas74}.
The resulting QFF is denoted as the CcCR QFF for the CBS energy, core
correlation correction, and scalar relativistic correction, respectively,
\citep{Fortenberry11HOCO}.  The augmented Dunning basis sets have been shown by
\cite{Skurski00} to be reliable for computations of anionic properties.  An
initial least-squares-fit of the CcCR energy points leads to a minor
transformation of the reference geometry such that the gradients are
identically zero.  This geometry and the resulting force constants are then
employed in the rovibrational computations.  All electronic structure
computations make use of the MOLPRO 2010.1 quantum chemical package
\citep{MOLPRO}, and all employ the Born-Oppenheimer approximation making the
QFFs identical for the isotopologues.

The QFF is fit from the 805 redundant total energy points with a sum of squared
residuals on the order of $3\times 10^{-17}$ a.u.$^2$  Cartesian derivatives
are then computed from the QFF with the INTDER program \citep{intder}.  From
these, the SPECTRO program \citep{spectro91} employs second-order vibrational
perturbation theory (VPT2) to generate the spectroscopic constants
\citep{Papousek82} and vibrational frequencies \citep{Mills72, Watson77}.
After transforming the force constants into the Morse-cosine coordinate system
so that the potential possesses proper limiting behavior \citep{Dateo94,
Fortenberry13Morse}, vibrational configuration interaction (VCI) computations
with the MULTIMODE program \citep{Carter98, Bowman03} also produce vibrational
frequencies.  The VCI computations make use of similar basis set configurations
as those utilized by \cite{Fortenberry12hococat, Fortenberry12HOCScat} in
similar quasi-linear tetra-atomic systems. 

\section{Discussion}

The force constants computed in this study are listed in Table \ref{fc1}.  The
CcCR geometrical parameters and spectroscopic constants are given in Table
\ref{StructHarm} for both $1\ ^1A'$ $l$-C$_3$H$^-$ and the deuterated
isotopologue.  The equilibrium dipole moment is computed with respect to the
center-of-mass with CCSD(T)/aug-cc-pV5Z to be 2.16 D.  The C$-$C$-$C
$R_{\alpha}$ vibrationally-averaged  bond angle is nearly collinear at
174.540$^{\circ}$ while the vibrationally-averaged $\angle$H$-$C$-$C is
109.491$^{\circ}$.  These values are in line with those computed by
\cite{Lakin01}.  As has been discussed by \cite{Fortenberry133DBS} for
C$_3$H$^-$, the C$_1$ carbon atom adjacent to the hydrogen atom shown in Figure
\ref{fig} is a carbene-type carbon containing a lone pair which leads to a
longer C$_1-$C$_2$ bond length compared to the shorter C$_2-$C$_3$ bond length.
Even though this result differs from the CCSD(T) results from \cite{Lakin01},
their reported CASSCF and HF results give bond lengths similar to ours leading
us to conclude that the CCSD(T) C$-$C bond lengths are mislabeled in the paper
by \cite{Lakin01}.  The vibrationally-averaged geometrical parameters change
slightly upon deuteration.  Similar bond angles of the heavy atoms have been
computed for the $trans$-HOCO$^+$, HOCS$^+$, and HSCO$^+$ systems
\citep{Fortenberry12hococat, Fortenberry12HOCScat} with very good agreement
present for known experimental data.

The most notable values in Table \ref{StructHarm} are the rotational constants
and the quartic centrifugal distortion ($D$-type) constants.  For $1\ ^1A'$
$l$-C$_3$H$^-$, the $B_0$ rotational constant is 11 339.66 MHz while $C_0$ is
11 087.35 MHz.  The equilibrium constants are slightly larger, but both sets
are in reasonable agreement with those computed by \cite{Lakin01}.  The
$D$-type constants have not been vibrationally-averaged, and $D_J$, most
prominently, is 4.954 kHz.

Direct comparison between these explicitly computed values and those deduced
from the Horsehead nebula PDR spectrum observed by \cite{Pety12} is not
possible since the isomer of C$_3$H$^-$ of interest here is not perfectly
linear.  \cite{Pety12} assume a linear structure in order to fit the effective
rotational constant, $B_{eff}$, and the effective centrifugal distortion
constant, $D_{eff}$ and use the second-order fitting equation,
\begin{equation}
\nu_{J+1\rightarrow J}=2B(J+1)-4D(J+1)^3,
\label{lin}
\end{equation}
to compute the affiliated rotational constants.  C$_3$H$^-$ is non-linear and
requires the following related equation from \cite{McCarthy97}:
\begin{equation}
\nu_{J+1\rightarrow J}=(B+C)(J+1)-\left\{4D_J+\frac{(B-C)^2}{c \left[ A-\frac{(B+C)}{2} \right] } \right\}(J+1)^3,
\label{nonlin}
\end{equation}
with the assumption that $K=0$ forcing $c=8$.  As such, we can set Equation
\ref{lin} equal to Equation \ref{nonlin}.  The $(J+1)$ term in Equation
\ref{nonlin} is equal to $2B_{eff}$, and the $(J+1)^3$ term in Equation
\ref{nonlin} is equal to $4D_{eff}$.  Using the CcCR computed $A_0$, $B_0$,
$C_0$, and $D_J$ values, where $D_J$ is the only equilibrium constant,
$B_{eff}$ is computed to be 11 213.51 MHz, and $D_{eff}$ is 8.795 kHz.  Hence,
direct comparison between the CcCR C$_3$H$^-$ derived effective rotational
constants and those obtained from the lines observed by \cite{Pety12} is
possible.

The second-order fit of the lines observed by \cite{Pety12} indicates that the
carrier must have a $B$-type constant that is very close to 11 244.9474 MHz and
a $D$-type quartic distortion constant that is around 7.652 kHz.  The $B_{eff}$
computed with the $A_0$, $B_0$, and $C_0$ rotational constants by the above
approach is very close, off by 31.44 MHz or 0.28\%.  This is roughly the same
difference between the observed $B$ and that of $l$-C$_3$H$^+$
\citep{Huang13C3H+}.  However, the 8.795 kHz $D_{eff}$ for $1\ ^1A'$
$l$-C$_3$H$^-$ is much closer to the 7.652 kHz $D$ derived from the lines
observed by \cite{Pety12} in the Horsehead nebula than the linear cation
\citep{Huang13C3H+}.  Even so, this $D_{eff}$ of 8.795 kHz differs from the
observation by 1.14 kHz or 14.93\%.


Table \ref{errors} provides some insight into the accuracies that can be
expected for calculated rotational constants of similar molecules.  Related
quasilinear molecules studied previously have all been cations.  Hence, within
Table \ref{errors}, the cation $B$ and $D$-type constants listed are more
correctly understood to be $B_{eff}$ and $D_{eff}$ as is the case for
C$_3$H$^-$ ($i.e.$ Equation \ref{nonlin} is used).  Calculation of the
vibrationally-averaged $B_{eff}$ values incorporate $B_0$ and $C_0$ while the
equilibrium $B_{eff}$ values incorporate $B_e$ and $C_e$.  Calculation of
$D_{eff}$ for each of the bent, quasilinear systems utilizes $A_0$, $B_0$, and
$C_0$ and the equilibrium $D_J$ value since vibrational averaging is not
available for the $D$-type constants.  The lone exception to this definition of
$D_{eff}$ is the C$_3$H$^-$ $D_{eff}$ computed with $A_e$, $B_e$, and $C_e$
given in the second line of Table \ref{errors}, which actually lowers the
C$_3$H$^-$ $D_{eff}$ value to 8.366 kHz, a difference of 0.714 kHz or 9.34\%
from that determined by \cite{Pety12}.  Finally, since all of the anions
observed in the ISM have been linear, directly comparable $B_0$, $B_e$, and
$D_e$ constants have been computed explicitly and are listed in Table
\ref{errors}. 

From Table \ref{errors}, the quasi-linear cations listed below C$_3$H$^-$ show
strong correlation between the computed $B_{eff}$ from the use of $B_0$ and
$C_0$ and the $B_{eff}$ derived from the various experiments.  Additionally,
the $D_{eff}$ values computed the same way with the equilibrium $D_J$ also show
good, albeit not as strong, correlation between theory and experiment.
Unfortunately, C$_3$H$^-$ has errors that are larger than this.  However, this
probably results from a combination of basis set incompleteness and
higher-order correlation effects.  Even though aug-cc-pVXZ basis sets used at
the CCSD(T) level of theory have been shown to be effective in the computation
of anionic properties \citep{Skurski00, Fortenberry11dbs}, higher-order
properties such as the $D$-type constants are more susceptible to even the
smallest errors.  This is clear for the cations as well, where the $D_{eff}$
values are not as accurate as the $B_{eff}$ values.

The known interstellar anions and the related C$_2$H$^-$ system, which has not
yet been detected in the ISM, are linear and have $B$ and $D$ computed
directly, either as B$_0$ or $B_e$ and $D_e$.  Note that the theoretical
rotational constants are not as accurate for the anions as they are for the
cations.  Most notably, the $B_e$/$B_0$ and $D_e$ values computed with a
CCSD(T)/aug-cc-pCV5Z cubic force field for C$_5$N$^-$ by \cite{Botschwina08}
are directly used in the identification of this anion in the ISM
\citep{Cernicharo08}.  As listed in Table \ref{errors}, agreement between
computed $B$ values and that necessary to match the observed rotational lines
actually worsens when $B_0$ is used instead of $B_e$, more than doubling the
percent error.  This is the same behavior as what is currently found for
C$_3$H$^-$.  Additionally, the $D_e$ percent error for C$_5$N$^-$, as compared
to observation, is 9.1\%, almost exactly what it is for C$_3$H$^-$ when using
the equilibrium rotational constants.  The force field employed by
\cite{Botschwina08} also includes core correlation like the CcCR QFF.  Hence,
the present rotational constants are in the same accuracy range for C$_3$H$^-$
as those used to detect C$_5$N$^-$ in the ISM.  Furthermore, the calculated
$D_e$ values compared to experiment for C$_6$H$^-$ and C$_8$H$^-$ actually have
a larger percent error than $D_{eff}$ for C$_3$N$^-$, C$_5$N$^-$, or even
C$_3$H$^-$.


Comparison of the sextic distortion constant, $H_{eff}$, is not as
straightforward.  There is a dearth of data on how the computation of this
value for anions compares to experiment.  $H_J$, which is an equilibrium value,
is not exactly $H_{eff}$, but they are probably related.  Even though
$H$ obtained by \cite{Pety12} is 560 mHz and $H_J$ for C$_3$H$^-$ is 3.344 mHz,
this is an order of magnitude closer agreement than this same $H$ compared to
the $H_e$ for $l$-C$_3$H$^+$, 0.375 mHz \citep{Huang13C3H+}.  Additionally, the
same basis set and correlation errors for anions that affect the calculation of
$D$ will be present for $H$.  As a result, we can only say here that as far as
$H$ is concerned for comparison to the lines observed in the Horsehead nebula
by \cite{Pety12}, $1\ ^1A'$ C$_3$H$^-$ is a better candidate than
$l$-C$_3$H$^+$.

Even though lower levels of theory have been used to reproduce rotational
constants of the detected, linear interstellar anions, C$_3$H$^-$ is the only
anion examined here that is not linear.  It is known that basis set effects can
be pronounced in the computation of bond angles in anions \citep{Lee85,
Huang09} where the average change in a bond angle computed with a standard
basis set and one augmented to include diffuse functions is around
1.0$^{\circ}$.  For example, the equilibrium $\angle$C$-$C$-$C in C$_3$H$^-$
from a simple CCSD(T)/cc-pVTZ QFF is 173.32$^{\circ}$, while this same angle is
174.20$^{\circ}$ with a CCSD(T)/aug-cc-pVTZ QFF.  In fact, the resulting
0.88$^{\circ}$ difference by simply adding diffuse functions to the standard
basis set is actually larger than the equilibrium $\angle$C$-$C$-$C difference
between the CCSD(T)/aug-cc-pVTZ QFF and that from the CcCR QFF, 0.38$^{\circ}$.
As a result, $B_{eff}$ for C$_3$H$^-$ is slower to converge with respect to the
basis set chosen relative to the linear anions.  This is made clear in that the
CCSD(T)/cc-pVTZ QFF vibrationally-averaged $B_{eff}$ is 11 056.74 MHz whereas
the corresponding CcCR $B_{eff}$ is 11 213.51 MHz, an increase of 156.77 MHz.
The linear anions are able to use lower level levels of theory in order to
approach the experimental rotational constants, but higher levels of theory are
required for the non-linear anion.  The fact that $B_{eff}$ computed with the
equilibrium rotational constants is closer to the $B$ derived from the
observations by \cite{Pety12} than $B_{eff}$ computed with the
vibrationally-averaged rotational constants is coincidental.  However, the
important point is that the C$_3$H$^-$ vibrationally-averaged $B_{eff}$
approaches the corresponding observed value as more accurate QFFs are employed,
and the remaining error is typical.

The harmonic and anharmonic vibrational frequencies for both $1\ ^1A'$
$l$-C$_3$H$^-$ and $l$-C$_3$D$^-$ are given in Table \ref{vptvci}.  Positive
anharmonicities are present in both isotopologues for the $\nu_5$
C$_1-$C$_2-$C$_3$ bending and the $\nu_6$ torsional modes.  VPT2 and VCI
produce fundamental vibrational frequencies from the CcCR QFF that are quite
consistent.  The largest deviation between the methods, 1.0 cm$^{-1}$, is found
for the $\nu_4$ H$-$C$_1-$C$_2$ bending mode.  Comparison of the C$_3$H$^-$
CcCR QFF anharmonic vibrational frequencies, whether using VPT2 or VCI, to
those computed by \cite{Lakin01} is roughly consistent for $\nu_1$-$\nu_4$.
The $\nu_5$ anharmonic frequencies differ by more than 50 cm$^{-1}$, though the
$\omega_5$ harmonic frequencies are very similar (i.e., the difference in the
$\nu_5$ fundamental frequency is mostly due to differences in the anharmonic
correction).  The torsional mode is nearly identical between the two studies,
though in this case the harmonic frequencies differ by more than 50 cm$^{-1}$.
It is hoped that the present QFF computations of the fundamental vibrational
frequencies provided here will assist in the characterization of this anion in
current and future studies of the ISM or simulated laboratory experiments at
infrared wavelengths in addition to studies in the sub-millimeter spectral
region.

\section{Astrochemical Considerations}

The lines observed by \cite{Pety12} are present in the Horsehead nebula PDR but
not in the dense core.  Typically, a PDR is defined in terms of shells starting
from the exterior shell dominated by an influx of far-ultraviolet (FUV)
photons.  In this region, the photons are most often absorbed by polycyclic
aromatic hydrocarbons (PAHs) and dust particles.  However, electrons are also
produced in these regions from various mechanisms involving the aforementioned
larger molecular particles as well as from interactions with atoms or small
molecules.  As the FUV flux is reduced from shielding resulting from the PAHs
and dust, the H$_2$ shell is formed.  Moving further in to the region, CO
begins to form, and, finally, O$_2$ formation is present in the dense core when
the photon shielding is high enough \citep{Tielens, Wolfire11}.  In fact, PDRs
are believed to be a major cache of the interstellar molecular abundance due to
the stability of the dense cores.


It could be assumed that such a large flux of high-energy photons in the outer
shells would remove any excess electron from an anion or even from many neutral
radicals.  However, this same process results in a veritable sea of elections
that could attach to neutrals and actually lead to the creation of anions even
in the Horsehead nebula PDR \citep{Millar07}.  Additionally, many anions are
known to be surprisingly stable \citep{Hammer03, Simons08, Simons11,
Fortenberry11dbs, Fortenberry112dbs, Fortenberry133DBS}, and electron
attachment rates are also believed to be quite high in these regions
\citep{Millar07}.  Several anions have also been shown to possess dipole-bound
excited states, or threshold resonances, which may play a significant role in
the creation and recreation of interstellar anions \citep{Guthe01, Carelli13}.
The mechanism of radiative attachement (RA), outlined by \cite{Carelli13} as
radiative stabilization, describes attachment of an electron to a neutral
speices, A, through creation of the excited electronic state of the resultant
anion:
\begin{equation}
\mathrm{A}+e^- \rightarrow \mathrm{[A^-]^*} \rightarrow \mathrm{A^-} + h\nu.
\label{RA}
\end{equation}
Relaxation can take place such that the electronic ground state of the anion
would be present \citep{Carelli13, Millar07, Herbst08}.


The dipole-bound (and only) excited state of a small anion typically should
function as the necessary excited state for RA.  Dipole-bound states are known
to exist for each of C$_4$H$^-$, C$_6$H$^-$, and C$_8$H$^-$ \citep{Pino02}.  In
order for such a state to be present, the dipole moment of the corresponding
neutral, a radical for these systems, must be on the order of 2 D or larger
\citep{Simons08, Simons11}.  For the $^2\Pi$ ground states of C$_6$H and
C$_8$H, the dipole moments are large enough to support a singlet dipole-bound
excited state.  C$_4$H, $^2\Sigma^+$ in its ground state \citep{Fortenberry10},
has a relatively small dipole moment at 0.8 D \citep{Graf01}.  Hence, in order
for C$_4$H$^-$ to form, the radical must either excite out of the weakly
dipolar $\tilde{X}\ ^2\Sigma^+$ state into the large-dipole $A\ ^2\Pi$ state
before undergoing RA, or it must form through another manner besides RA.  As
discussed by \cite{Gupta07} and \cite{McCarthy08C3N-}, the need for radical
excitation followed by RA could explain the very low $[$C$_4$H$^-/$C$_4$H$]$
ratio observed towards various interstellar objects \citep{Agundez08,
Cordiner13}.  Even though these two states of C$_4$H are ``nearly degenerate"
\citep{Taylor98}, some additional energy is required to populate the $A\ ^2\Pi$
state, which, in turn, lowers the probability of electron attachment.
Furthermore, C$_2$H is also $^2\Sigma^+$ in its ground state, but the
excitation energy into the large-dipole $A\ ^2\Pi$ state is more than double
its counterpart in C$_4$H \citep{Fortenberry10}, which may shed light on the
even lower $[$C$_2$H$^-/$C$_2$H$]$ ratio proposed by \cite{Agundez08}.
C$_6$H$^-$ and C$_8$H$^-$ could be present in the Horsehead nebula PDR, as has
been suggested from observations and modeling by \cite{Agundez08}, but these
longer anions may only be accessible from their dipole-bound excited states.
However, C$_3$H$^-$ has more than just a dipole-bound excited state.


A few, rare anions possess valence excited electronic states between the
dipole-bound state and the ground electronic state \citep{Fortenberry11dbs,
Fortenberry112dbs}.  As mentioned in the Introduction, $1\ ^1A'$ C$_3$H$^-$ is,
thus far, the only anion composed solely of first-row atoms (and hydrogen) to
possess a valence singlet excited state \citep{Fortenberry133DBS}.  The
presence of two excited electronic states with the same spin multiplicity should
increase the production of C$_3$H$^-$ since multiple relaxation pathways exist.
Beginning from the dipole-bound state, the excited anion can relax within the
RA mechanism to the ground electronic state either directly or via the valence
excited state first.  Enough C$_3$H$^-$ may then exist in a steady state to
counterbalance the destructive photons present in this region.




If the Horsehead nebula PDR abundances of $l$-C$_3$H$^+$ from \cite{Pety12} can
be inferred actually to be $1\ ^1A'$ C$_3$H$^-$, the $[$C$_3$H$^-/$C$_3$H$]$
ratio could be as high as 0.30 in the Horsehead nebula PDR.  This is not as
high as the upper limit proposed for $[$C$_6$H$^-/$C$_6$H$]$ at 8.9, but it is
an order of magnitude larger than $[$C$_4$H$^-/$C$_4$H$]$ \citep{Agundez08} as
can be expected since the ground electronic state of C$_3$H is strongly dipolar
and that of C$_4$H is not.  The amount of C$_3$H$^-$ should decrease as the
observations move towards the dense molecular core due to the higher reactivity
of this anion.  The reaction cross-section of anions is much larger than in
neutrals \citep{Eichelberger07}, and C$_3$H$^-$ could go through various
destructive processes \citep{Millar07, Larsson12} as the molecular density
increases.  Alternatively, this anion could exist within the observed sightline
but on the outer edge of the PDR where the photon flux is small enough for a
measurable population to be stable.  In this region a longer path length of
such material is also present away from the high $A_V$ dense core.  Either way,
the existence of $1\ ^1A'$ C$_3$H$^-$ in the Horsehead nebula PDR is feasible.

\section{Conclusions}

Since the link between $l$-C$_3$H$^+$ and the lines observed in the Horsehead
nebula PDR by \cite{Pety12} has recently been strongly questioned by
\cite{Huang13C3H+}, another viable candidate is necessary.  The rotational
lines seem to require a closed-shell quasi-linear structure composed of three
carbon atoms along with a hydrogen atom.  $1\ ^1A'$ C$_3$H$^-$ appears to be
the most likely candidate.  Here, the CcCR QFF has determined a $B_{eff}$ for
this anion to be in error by 0.28\% from that required to fit the observed
lines.  The use of the equilibrium rotational constants fortuitously lowers the
error to 0.11\%.  However, the error reduction and error magnitudes themselves
are in line with the computed C$_5$N$^-$ rotational constants used in its
interstellar detection.  Additionally, the discrepancy between the $A_e$,
$B_e$, and $C_e$ computed C$_3$H$^-$ $D_{eff}$ and the $D_{eff}$ deduced from
the observed interstellar rotational lines is similar to the $D_e$ errors for
C$_4$H$^-$, C$_3$N$^-$, and C$_5$N$^-$ and less than that of C$_6$H$^-$, which
are all reported for CCSD(T) computations, $i.e.$ similar levels of theory.
Hence, the consistency of the errors for C$_3$H$^-$ with other anions
previously detected in the ISM coupled with its matching the required spectral
criteria and the rationale for its existence involving its valence and
dipole-bound excited states, make this anion the strongest candidate carrier
for the Horsehead nebula PDR lines and, potentially, the seventh and most
recent anion detected in the ISM.  It would also be the first detected
interstellar odd-numbered carbon monohydrogen chain anion.


\section{Acknowledgements}

\begin{figure}[h]
\caption{CcCR equilibrium geometry of $1\ ^1A'$ $l$-C$_3$H$^-$.}
\includegraphics[width = 6.0 in]{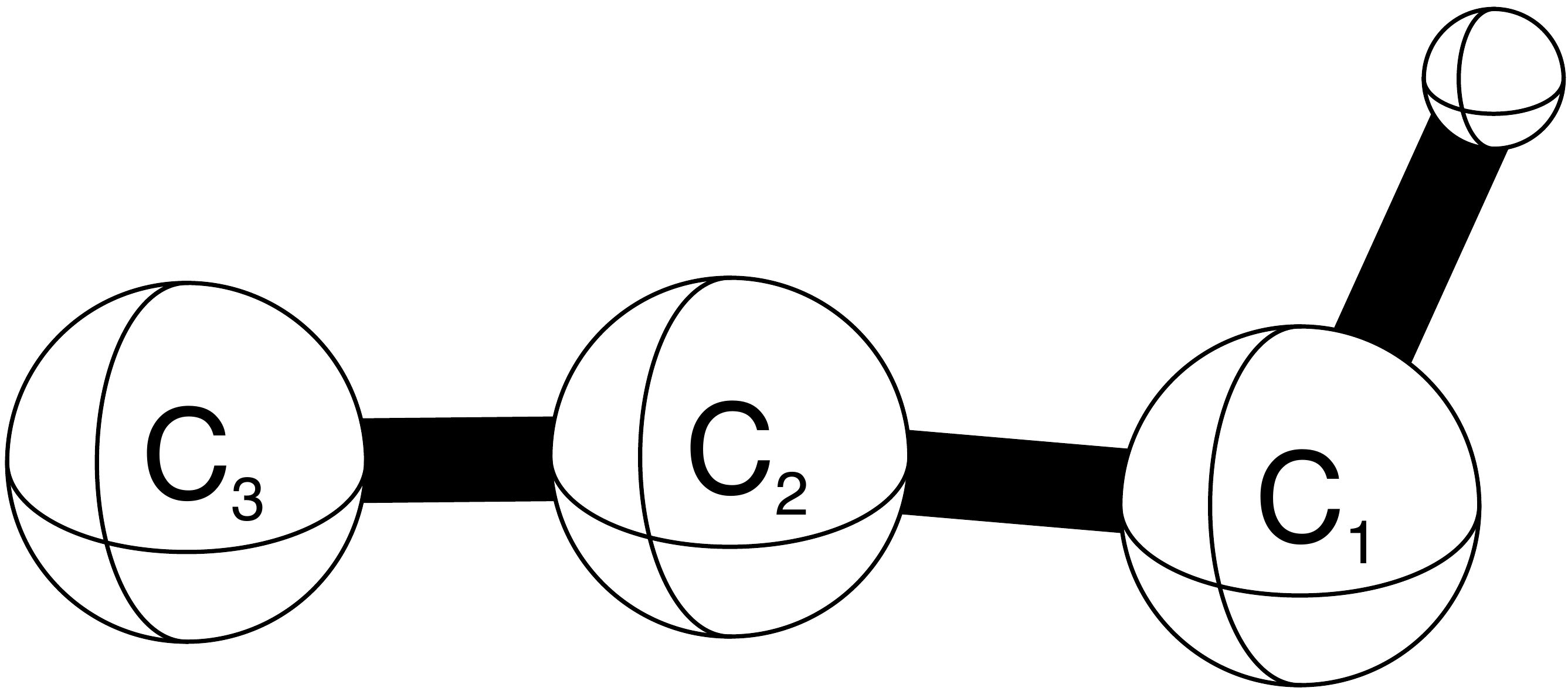}
\label{fig}
\end{figure}

RCF is currently supported on a NASA Postdoctoral Program Fellowship
administered by Oak Ridge Associated Universities.  NASA/SETI Institute
Cooperative Agreement NNX12AG96A has funded the work undertaken by XH.  Support
from NASA's Laboratory Astrophysics `Carbon in the Galaxy' Consortium Grant
(NNH10ZDA001N) is gratefully acknowledged.  The U.S.~National Science
Foundation (NSF) Multi-User Chemistry Research Instrumentation and Facility
(CRIF:MU) award CHE-0741927 provided the computational hardware, and award
NSF-1058420 has supported TDC.  The CheMVP program was used to create
Fig.~\ref{fig}.  The authors would also like to acknowledge many others for
their contributions to our astronomical understanding of this subject.  These
include, most notably: Dr.~Michael C.~McCarthy of the Harvard-Smithsonian
Center for Astrophysics, Dr.~Naseem Rangwala of the University of Colorado,
Dr.~Lou Allamandola of the NASA Ames Research Center, and Dr.~Christiaan
Boersma of the NASA Ames Research Center and San Jose State Univeristy.

\bibliographystyle{apj}

\newpage

\renewcommand{\baselinestretch}{1}

\begingroup
\begin{table}[h]

\caption{The simple-internal CcCR QFF Quadratic, Cubic, and Quartic Force
Constants (in mdyn/\AA$^n$$\cdot$rad$^m$)$^a$ for $l$-C$_3$H$^-$.}

\label{fc1}

\centering
\small

\begin{tabular}{c r c r c r c r c r}
\hline

\hspace{0.1in}$F_{11}$\hspace{0.1in} & 10.191 889 &\hspace{0.1in}
$F_{431}$\hspace{0.1in} & 0.0711 & \hspace{0.1in} $F_{1111}$\hspace{0.1in} &
318.24 & \hspace{0.1in} $F_{4432}$\hspace{0.1in} & 0.24 &\hspace{0.1in}
$F_{5531}$\hspace{0.1in} &  0.12\\
$F_{21}$ &  0.841 962 & $F_{432}$ & -0.4022 & $F_{2111}$ &   0.44 & $F_{4433}$ &  0.44 & $F_{5532}$ &  0.22 \\  
$F_{22}$ &  7.312 189 & $F_{433}$ & -0.0735 & $F_{2211}$ &  -1.98 & $F_{4441}$ &  0.45 & $F_{5533}$ & -0.34 \\  
$F_{31}$ &  0.068 029 & $F_{441}$ & -0.5015 & $F_{2221}$ &   5.28 & $F_{4442}$ & -0.34 & $F_{5541}$ & -0.01 \\  
$F_{32}$ & -0.006 196 & $F_{442}$ &  0.1723 & $F_{2222}$ & 220.01 & $F_{4443}$ &  0.56 & $F_{5542}$ &  0.07 \\  
$F_{33}$ &  4.558 746 & $F_{443}$ & -0.0586 & $F_{3111}$ &   0.16 & $F_{4444}$ & -0.81 & $F_{5543}$ & -0.17 \\  
$F_{41}$ & -0.066 879 & $F_{444}$ & -0.7769 & $F_{3211}$ &  -0.08 & $F_{5111}$ &  0.06 & $F_{5544}$ &  0.51 \\  
$F_{42}$ &  0.515 214 & $F_{511}$ & -0.0809 & $F_{3221}$ &   0.80 & $F_{5211}$ &  0.10 & $F_{5551}$ &  0.07 \\  
$F_{43}$ &  0.217 498 & $F_{521}$ & -0.0018 & $F_{3222}$ &  -1.41 & $F_{5221}$ & -0.39 & $F_{5552}$ &  0.18 \\  
$F_{44}$ &  0.650 100 & $F_{522}$ & -0.3714 & $F_{3311}$ &   0.72 & $F_{5222}$ &  0.80 & $F_{5553}$ &  0.12 \\  
$F_{51}$ &  0.070 974 & $F_{531}$ & -0.0597 & $F_{3321}$ &  -0.80 & $F_{5311}$ &  0.09 & $F_{5554}$ & -0.12 \\  
$F_{52}$ &  0.081 130 & $F_{532}$ & -0.2190 & $F_{3322}$ &  -0.54 & $F_{5321}$ &  0.21 & $F_{5555}$ &  1.95 \\  
$F_{53}$ &  0.069 481 & $F_{533}$ & -0.0030 & $F_{3331}$ &  -0.82 & $F_{5322}$ &  0.54 & $F_{6611}$ &  0.14 \\  
$F_{54}$ &  0.064 059 & $F_{541}$ & -0.0774 & $F_{3332}$ &   0.46 & $F_{5331}$ & -0.04 & $F_{6621}$ & -0.49 \\  
$F_{55}$ &  0.404 485 & $F_{542}$ &  0.0376 & $F_{3333}$ & 145.05 & $F_{5332}$ & -0.07 & $F_{6622}$ &  0.64 \\  
$F_{66}$ &  0.168 044 & $F_{543}$ & -0.0694 & $F_{4111}$ &  -0.17 & $F_{5333}$ & -0.46 & $F_{6631}$ &  0.03 \\  
$F_{111}$ & -64.7214 &  $F_{544}$ & -0.1425 & $F_{4211}$ &  -0.02 & $F_{5411}$ &  0.14 & $F_{6632}$ &  0.06 \\  
$F_{211}$ &   0.5759 &  $F_{551}$ & -0.4284 & $F_{4221}$ &  -0.27 & $F_{5421}$ &  0.12 & $F_{6633}$ & -0.14 \\  
$F_{221}$ &  -3.2972 &  $F_{552}$ & -0.9210 & $F_{4222}$ &  -0.32 & $F_{5422}$ & -0.49 & $F_{6641}$ & -0.07 \\  
$F_{222}$ & -43.3783 &  $F_{553}$ & -0.0900 & $F_{4311}$ &   0.15 & $F_{5431}$ & -0.05 & $F_{6642}$ & -0.04 \\  
$F_{311}$ &   0.0659 &  $F_{554}$ &  0.0071 & $F_{4321}$ &  -0.03 & $F_{5432}$ &  0.31 & $F_{6643}$ & -0.10 \\  
$F_{321}$ &  -0.3042 &  $F_{555}$ & -0.1839 & $F_{4322}$ &  -0.46 & $F_{5433}$ & -0.09 & $F_{6644}$ &  0.08 \\  
$F_{322}$ &  -0.0840 &  $F_{661}$ & -0.1710 & $F_{4331}$ &  -0.06 & $F_{5441}$ &  0.01 & $F_{6651}$ & -0.04 \\  
$F_{331}$ &   0.1287 &  $F_{662}$ & -0.3467 & $F_{4332}$ &  -0.31 & $F_{5442}$ &  0.14 & $F_{6652}$ &  0.02 \\  
$F_{332}$ &   0.2601 &  $F_{663}$ & -0.0476 & $F_{4333}$ &  -1.42 & $F_{5443}$ &  0.24 & $F_{6653}$ & -0.03 \\  
$F_{333}$ & -28.7819 &  $F_{664}$ &  0.0133 & $F_{4411}$ &  -0.63 & $F_{5444}$ & -0.06 & $F_{6654}$ & -0.10 \\  
$F_{411}$ &  -0.2017 &  $F_{665}$ & -0.0708 & $F_{4421}$ &   1.45 & $F_{5511}$ &  0.57 & $F_{6655}$ &  0.23 \\  
$F_{421}$ &   0.3282 &            &         & $F_{4422}$ &  -1.71 & $F_{5521}$ &  0.14 & $F_{6666}$ &  0.86 \\  
$F_{422}$ &  -0.6518 &            &         & $F_{4431}$ &  -0.17 & $F_{5522}$ &  1.56 \\ 
\hline
                      
\end{tabular}

$^a$1 mdyn $=$ $10^{-8}$ N; $n$ and $m$ are exponents corresponding to the number
of units from the type of modes present in the specific force constant.

\end{table}
\endgroup
\renewcommand{\baselinestretch}{2}

\renewcommand{\baselinestretch}{1}
\begingroup
\begin{table}[h]

\caption{The Zero-Point ($R_{\alpha}$ vibrationally-averaged) and Equilibrium
Structures, Rotational Constants, CCSD(T)/aug-cc-pV5Z Dipole Moment,
Vibration-Rotation Interaction Constants, and Quartic and Sextic Distortion
Constants of $1\ ^1A'$ $l$-C$_3$H$^-$ and the deuterated form with the CcCR QFF.}

\label{StructHarm}

\centering

\tiny

\begin{tabular}{l | r r r} 
\hline\hline

                              & C$_3$H$^-$ & Previous$^a$ & C$_3$D$^-$ \\
\hline
r$_0$(C$_1-$H)                & 1.119 438 \AA &  & 1.116 446 \AA  \\
r$_0$(C$_1-$C$_2$)            & 1.351 595 \AA &  & 1.351 753 \AA  \\
r$_0$(C$_2-$C$_3$)            & 1.282 845 \AA &  & 1.282 620 \AA  \\
$\angle_0$(H$-$C$_1-$C$_2$)   & 109.491$^{\circ}$ &  & 109.530$^{\circ}$ \\
$\angle_0$(C$_1-$C$_2-$C$_3$) & 174.540$^{\circ}$ &  & 174.643$^{\circ}$ \\
$A_0$                         & 529 134.2 MHz &  & 295 539.6 MHz \\
$B_0$                         & 11 339.66 MHz &  & 10 626.03 MHz \\
$C_0$                         & 11 087.35 MHz &  & 10 238.74 MHz \\
$D_J$                         & 4.954 kHz     &  &  4.544 kHz    \\
$D_{JK}$                      & 0.702 MHz     &  &  0.316 MHz    \\
$D_K$                         & 217.543 MHz   &  &  94.897 MHz   \\
$d_1$                         & -0.112 kHz    &  &  -0.253 kHz   \\
$d_2$                         & -0.023 kHz    &  &  -0.052 kHz   \\
$H_J$                         & 3.344 mHz     &  &  16.516 mHz   \\
$H_{JK}$                      & 3.221 Hz      &  &  2.151 Hz     \\
$H_{KJ}$                      & -3.229 kHz    &  & -0.745 kHz    \\
$H_K$                         & 358.867 kHz   &  &  90.731 kHz   \\
$H_1$                         &  0.132 mHz    &  &  0.634 mHz    \\
$H_2$                         &  0.203 mHz    &  &  0.612 mHz    \\
$H_3$                         &  0.037 mHz    &  &  0.133 mHz    \\
$\tau_{aaaa}$                 & -873.001 MHz  &  &-380.872 MHz   \\
$\tau_{bbbb}$                 &  -0.021 MHz   &  &  -0.021 MHz   \\
$\tau_{cccc}$                 &  -0.019 MHz   &  &  -0.017 MHz   \\
$\tau_{aabb}$                 &  -2.766 MHz   &  &  -1.619 MHz   \\
$\tau_{aacc}$                 &  -0.081 MHz   &  &   0.319 MHz   \\
$\tau_{bbcc}$                 &  -0.020 MHz   &  &  -0.018 MHz   \\
$\Phi_{aaa}$                 & 355 640.661 Hz &  & 89 988.504 Hz \\
$\Phi_{bbb}$                  &    0.001 Hz   &  &   0.004 Hz    \\
$\Phi_{ccc}$                  &    0.000 Hz   &  &   0.001 Hz    \\
$\Phi_{aab}$                  &  390.158 Hz   &  & 703.204 Hz    \\
$\Phi_{abb}$                  &    4.265 Hz   &  &   3.112 Hz    \\
$\Phi_{aac}$                  &-3 614.354 Hz  &  &-1 445.590 Hz  \\
$\Phi_{bbc}$                  &    0.000 Hz   &  &   0.001 Hz    \\
$\Phi_{acc}$                  &   -0.271 Hz   &  &   0.151 Hz    \\
$\Phi_{bcc}$                  &    0.001 Hz   &  &   0.002 Hz    \\
$\Phi_{abc}$                  &    4.570 Hz   &  &   3.618 Hz    \\
$\alpha^A$ 1                  & 27 922.5 MHz  &  & 11 662.9 MHz  \\
$\alpha^A$ 2                  &   -725.5 MHz  &  &  -917.5 MHz   \\
$\alpha^A$ 3                  &    484.8 MHz  &  &   170.2 MHz   \\
$\alpha^A$ 4                  &-35 092.1 MHz  &  &-16 226.3 MHz  \\
$\alpha^A$ 5                  & -3 103.1 MHz  &  &-4 597.5 MHz   \\
$\alpha^A$ 6                  & 12 333.4 MHz  &  & 9 042.1 MHz   \\
$\alpha^B$ 1                  &   4.2 MHz     &  &   6.9 MHz     \\
$\alpha^B$ 2                  &  83.5 MHz     &  &  77.2 MHz     \\
$\alpha^B$ 3                  &  45.1 MHz     &  &  40.3 MHz     \\
$\alpha^B$ 4                  & -12.0 MHz     &  &  -8.4 MHz     \\
$\alpha^B$ 5                  & -47.1 MHz     &  & -48.4 MHz     \\
$\alpha^B$ 6                  & -48.6 MHz     &  & -45.9 MHz     \\
$\alpha^C$ 1                  &  14.8 MHz     &  &  18.3 MHz     \\
$\alpha^C$ 2                  &  78.6 MHz     &  &  70.1 MHz     \\
$\alpha^C$ 3                  &  38.4 MHz     &  &  39.3 MHz     \\
$\alpha^C$ 4                  &  16.0 MHz     &  &  12.9 MHz     \\
$\alpha^C$ 5                  & -16.1 MHz     &  & -15.2 MHz     \\
$\alpha^C$ 6                  & -78.5 MHz     &  & -69.8 MHz     \\
\hline                                                    
r$_e$(C$_1-$H)$^b$            & 1.106 939 \AA & 1.110 \AA & --  \\
r$_e$(C$_1-$C$_2$)            & 1.349 832 \AA & 1.289 \AA & --  \\
r$_e$(C$_2-$C$_3$)            & 1.281 900 \AA & 1.363 \AA & --  \\
$\angle_e$(H$-$C$_1-$C$_2$)   & 109.529$^{\circ}$ & 109.2$^{\circ}$ & --  \\
$\angle_e$(C$_2-$C$_3-$C$_4$) & 174.571$^{\circ}$ & 171.2$^{\circ}$ & --  \\
$A_e$                         & 530 044.3 MHz & 524.5 GHz & 295 106.5 MHz \\
$B_e$                         & 11 352.05 MHz &  11.2 GHz & 10 636.73 MHz \\
$C_e$                         & 11 114.02 MHz &  10.9 GHz & 10 266.68 MHz \\
$\mu$$^c$                     & 2.16 D        & --  & --  \\
$\mu_x$                       & 1.63 D        & --  & --  \\
$\mu_y$                       & 1.41 D        & --  & --  \\

\hline
\end{tabular}

$^a$CCSD(T)/aug-cc-pVQZ QFF results from \cite{Lakin01}.

$^b$The equilibrium geometries are identical among isotopologues from the use
of the Born-Oppenheimer approximation.\\

$^c$The C$_3$H$^-$ coordinates (in \AA\ with the center-of-mass at the origin)
used to generate Born-Oppenheimer dipole moment components are: H, 1.733414,
-0.910473, 0.000000; C$_1$,  1.276456,  0.098036, 0.000000; C$_2$, -0.069613,
-0.016965, 0.000000; C$_3$, -1.352424, -0.004605, 0.000000.\\

\end{table}
\endgroup

\begingroup
\begin{table}[h]

\caption{Errors in the computation of $B$ (in MHz) and $D$ (in kHz) for linear
molecules and $B_{eff}$ (in MHz) and $D_{eff}$ (in kHz) for quasilinear
molecules.}

\centering
\begin{tabular}{c | c | l l l | l l l}
\hline\hline
\label{errors}

 & Theoretical & \multicolumn{3}{c|}{$B$/$B_{eff}$} & \multicolumn{3}{c}{$D$/$D_{eff}$} \\
Molecule & $B_0$ or $B_e$ & Experiment & Theory & \% Error & Experiment & Theory & \% Error \\

\hline
C$_3$H$^-$$^a$ & Equilibrium & 11244.9474 & 11233.04 & 0.11\% & 7.652 & 8.366 &
9.3\% \\
   & Vib.-avg. & 11244.9474 & 11213.51 & 0.28\% & 7.652 & 8.795 & 14.9\% \\
\hline
HSCO$^+$$^b$ & Vib.-avg. &  5636.866   &  5637.60 & 0.01\% & 3.1      & 3.116 & 0.5\% \\
HOCO$^+$$^c$ & Vib.-avg. & 10691.58265 & 10705.44 & 0.13\% & 4.580576 & 4.511 & 1.5\% \\
NNOH$^+$$^d$ & Vib.-avg. & 11192.9214  & 11198.57 & 0.05\% & 7.764972 & 7.604 & 2.1\% \\
HOCS$^+$$^e$ & Vib.-avg. &  5726.66011 &  5730.22 & 0.06\% & 1.064    & 1.107 & 4.0\% \\
\hline
C$_2$H$^-$$^f$ & Equilibrium & 41639.20 & 41781.0 & 0.34\% & 0.09697 & 0.0946 &
2.4\% \\
  & Vib.-avg. & 41639.20   & 41614.0 & 0.06\% & \\
C$_4$H$^-$$^g$ & Equilibrium & 4654.9449  & 4625.6546 & 0.63\% & 0.5875  & 0.55
& 6.4\% \\
  & Vib.-avg. & 4654.9449  & 4653.9 & 0.02\% & \\
C$_6$H$^-$$^h$ & Vib.-avg. & 1376.86298 & 1376.9 & 0.00\% & 0.03235 & 0.0270 & 16.5\% \\
C$_8$H$^-$$^i$ & Vib.-avg. & 583.30404  & 583.2  & 0.02\% & 0.0042  & 0.0033 & 16.7\% \\
CN$^-$$^j$ & Equilibrium & 56132.7562 & 56152  & 0.03\% & 186.427 & 185 & 0.8\% \\
     & Zero-Point & 56132.7562  & 56126.5 & 0.01\% & \\
C$_3$N$^-$$^k$ & Equilibrium & 4851.62183 & 4850 & 0.03\% & 0.68592 & 0.628 & 8.4\% \\
C$_5$N$^-$$^l$ & Equilibrium & 1388.860 & 1387.8 & 0.08\% & 0.033 & 0.0300 & 9.1\% \\
               & Vib.-avg. & 1388.860 & 1386.2 & 0.19\% & \\

\hline\hline
\end{tabular}

\raggedright
\small

$^a$This work with the observed lines described by \cite{Pety12}.\\

$^b$CcCR QFF data \cite{Fortenberry12HOCScat} and experimental data from
\cite{Ohshima96}.\\

$^c$CcCR QFF data from \cite{Fortenberry12hococat} and experimental data from
\cite{Bogey88}.\\

$^d$CcCR QFF data from \cite{Huang13NNOH+}, experimental $B_{eff}$ from
\cite{McCarthy10}, and experimental $D_{eff}$ computed from the constants given
in \cite{Bogey88NNOH}.\\

$^e$CcCR QFF data from \cite{Fortenberry12HOCScat} and experimental data from
\cite{McCarthy07}.\\

$^f$CcCR QFF data from \cite{Huang09} and experimental data from
\cite{Brunken07}.\\

$^g$B$_0$ from the CCSD(T)/cc-pVTZ $B_e$ corrected for vibrational averaging
with CCSD(T)/cc-pVDZ; CCSD(T)/cc-pVDZ $D_e$; and experimental data are from
\cite{Gupta07}.  The RCCSD(T)/aug-cc-pVQZ $B_e$ is from \cite{Senent10}.

$^h$CCSD(T)/cc-pVTZ $B_e$ corrected for vibrational averaging with
CCSD(T)/cc-pVDZ, CCSD(T)/cc-pVDZ $D_e$, and experimental data from
\cite{McCarthy06}.\\

$^i$CCSD(T)/cc-pVTZ $B_e$ corrected for vibrational averaging with SCF/DZP,
SCF/DZP $D_e$, and experimental data from \cite{Gupta07}.\\

$^j$CCSD(T)/aug-cc-pCV5Z $B_e$, CCSD(T)/aug-cc-pCVQZ $D_e$, and experimental data
from \cite{Gottlieb07} with CCSD(T)/MTcc $B_0$ from \cite{Lee99}.\\

$^k$CCSD(T)/aug-cc-pCV5Z $B_e$ and $D_e$ from \cite{Kolos08} ($\Delta B_0$ is
reported to be 0.606 MHz giving a \% error of about 0.02\%) and experimental data
from \cite{Thaddeus08C3N-}. 

$^l$CCSD(T)/aug-cc-pCV5Z $B_e$ and $B_0$ with CCSD(T)/aug-cc-pVQZ $D_e$ from
\cite{Botschwina08} with experimental data from \cite{Cernicharo08}.\\

\end{table}
\endgroup

\begingroup
\begin{table}[h]

\caption{The C$_3$H$^-$ and C$_3$D$^-$ CcCR QFF harmonic, VCI, and VPT2
fundamental vibrational frequencies in cm$^{-1}$.}

\centering
\scriptsize
\begin{tabular}{c l | c c c | c c | c c c}
\hline\hline
\label{vptvci}

 & & \multicolumn{3}{c|}{C$_3$H$^-$} & \multicolumn{2}{c|}{Previous$^a$
C$_3$H$^-$} & \multicolumn{3}{c}{C$_3$D$^-$} \\

\hspace{-0.12in} Mode \hspace{-0.12in} & \multicolumn{1}{c|}{Description} &
Harmonic & VCI & VPT2 & Harmonic & Anharm. & Harmonic &
VCI & VPT2\\ 


\hline

$\nu_1(a')$ & C$_1-$H stretch        & 2881.9 &	2714.4 & 2713.9 & 2863 & 2723 & 2122.9 & 2036.4 & 2035.5 \\
$\nu_2(a')$ & C$_2-$C$_3$ stretch    & 1843.9 &	1804.3 & 1804.4 & 1831 & 1828 & 1832.9 & 1796.5 & 1796.5 \\
$\nu_3(a')$ & C$_1-$C$_2$ stretch    & 1117.1 &	1108.0 & 1107.9 & 1091 & 1120 & 1112.0 & 1100.9 & 1101.0 \\
$\nu_4(a')$ & H$-$C$_1-$C$_2$ bend   & 1037.8 &	1012.1 & 1011.1 & 1002 & 1022 & 817.0  & 803.8  & 802.7  \\
$\nu_5(a')$ & C$_1-$C$_2-$C$_3$ bend & 406.7  & 419.4  & 418.9  & 393  & 368  & 379.1  & 382.4  & 381.9  \\
$\nu_6(a'')$ & torsion               & 281.0  & 296.8  & 296.1  & 349  & 297  & 278.9  & 286.7  & 286.1  \\

\hline\hline
\end{tabular}

$^a$CCSD(T)/aug-cc-pVQZ QFF results from \cite{Lakin01}.

\end{table}
\endgroup

\end{document}